\documentstyle[12pt]{article}

\input{tcilatex}

\begin{document}

\author{C. Bizdadea\thanks{%
e-mail addresses: bizdadea@central.ucv.ro and bizdadea@hotmail.com}, E. M.\
Cioroianu, S. O. Saliu\thanks{%
e-mail addresses: osaliu@central.ucv.ro and odile\_saliu@hotmail.com} \\
Department of Physics, University of Craiova\\
13 A. I. Cuza Str., Craiova RO-1100, Romania}
\title{Freedman-Townsend vertex from Hamiltonian BRST cohomology}
\maketitle

\begin{abstract}
Consistent interactions among a set of two-form gauge fields in four
dimensions are derived along a Hamiltonian cohomological procedure. It is
shown that the deformation of the BRST charge and BRST-invariant Hamiltonian
for the free model leads to the Freedman-Townsend interaction vertex. The
resulting interaction deforms both the gauge transformations and
reducibility relations, but not the algebra of gauge transformations.

\end{abstract}

\section{Introduction}

The cohomological understanding of the antifield-BRST symmetry \cite{2}--%
\cite{3} was proved to be a useful tool for constructing consistent
interactions in gauge theories \cite{15}--\cite{18}. Among the models of
great interest in theoretical physics that have been inferred along the
deformation of the master equation, we mention the Yang-Mills theory \cite
{19}, the Freedman-Townsend model \cite{20}, and the Chapline-Manton model 
\cite{21}. Also, it is important to notice the deformation results connected
to Einstein's gravity theory \cite{22} and four- and eleven-dimensional
supergravity \cite{23}.

Recently, first-order consistent interactions among exterior $p$-forms have
been approached in \cite{24} also by means of the antifield-BRST
formulation. On the one hand, models with $p$-form gauge fields
(antisymmetric tensor fields of various ranks) play an important role in
string and superstring theory, supergravity and the gauge theory of gravity 
\cite{26}--\cite{29}. The study of theoretical models with gauge
antisymmetric tensor fields give an example of so-called `topological field
theory' and lead to the appearance of topological invariants, being thus in
close relation to space-time topology, hence with lower dimensional quantum
gravity \cite{27}. In the meantime, antisymmetric tensor fields of various
orders are included within the supergravity multiplets of many supergravity
theories \cite{28}, especially in $10$ or $11$ dimensions. The construction
of `dual' Lagrangians involving $p$-forms is naturally involved with General
Relativity and supergravity in order to render manifest the $SL\left( 2,{\bf %
R}\right) $ symmetry group of stationary solutions of Einstein's vacuum
equation, respectively to reveal some subtleties of `exact solutions' for
supergravity \cite{29}. On the other hand, the Hamiltonian version of BRST
formalism \cite{1}, \cite{3} appears to be the most natural setting for
implementing the BRST symmetry in quantum mechanics \cite{3} (Chapter 14),
as well as for establishing a proper connection with canonical quantization
formalisms, like, for instance, the reduced phase-space or Dirac
quantization procedures \cite{30}. These considerations motivate the
necessity of a Hamiltonian BRST approach to consistent interactions that can
be added among a system of exterior $p$-forms. To our knowledge, this
problem has not been analyzed until now.

In this paper we investigate the consistent Hamiltonian interactions that
can be introduced among a set of abelian two-form gauge fields in four
dimensions. Our programme is the following. Initially, we show that the
Hamiltonian problem of introducing consistent interactions among fields with
gauge freedom can be reformulated as a deformation problem of the BRST
charge and BRST-invariant Hamiltonian of a given ``free'' theory, and derive
the main equations describing these two types of deformations, which turn
out to involve the ``free'' Hamiltonian BRST differential. Subsequently, we
start with the action of a set of abelian two-forms in four dimensions (in
first-order form), and construct its corresponding BRST charge and
BRST-invariant Hamiltonian. The BRST symmetry of this free model splits as
the sum between the Koszul-Tate differential and the exterior derivative
along the gauge orbits. Next, we solve the main equations that govern the
Hamiltonian deformation procedure in the case of the model under study
taking into account the BRST cohomology of the free theory. As a result of
this cohomological approach, we find the BRST charge and BRST-invariant
Hamiltonian of the deformed model. Relying on these deformed quantities, we
then identify the deformed Hamiltonian theory by analyzing its first-class
constraints, first-class Hamiltonian and also the corresponding gauge
algebra plus reducibility relations. The resulting system is nothing but the
non-abelian Freedman-Townsend model in four dimensions \cite{32}.

\section{Hamiltonian deformation equations}

We consider a dynamical ``free'' theory, described by the canonical
variables $z^{A}$, subject to the first-class constraints 
\begin{equation}
G_{a_{0}}\left( z^{A}\right) \approx 0,\;a_{0}=1,\ldots ,M_{0},  \label{ft1}
\end{equation}
that can in principle be reducible. For definiteness, we take all the
canonical variables to be bosonic, but our analysis can be extended to
fermions modulo introducing some appropriate sign factors. It is well known
that a constrained Hamiltonian system can be described by the action 
\begin{equation}
S_{0}\left[ z^{A},u^{a_{0}}\right] =\int\limits_{t_{1}}^{t_{2}}dt\left(
a_{A}\left( z\right) \dot{z}^{A}-H_{0}-u^{a_{0}}G_{a_{0}}\right) ,
\label{ft2}
\end{equation}
where $H_{0}$ is the first-class Hamiltonian, $u^{a_{0}}$ stands for the
Lagrange multipliers, and $a_{A}\left( z\right) $ is the one-form potential
that induces a symplectic two-form $\omega _{AB}$, whose inverse $\omega
^{AB}$ defines the fundamental Dirac brackets $\left[ z^{A},z^{B}\right]
^{*}=\omega ^{AB}$. The Hamiltonian gauge algebra reads as 
\begin{equation}
\left[ G_{a_{0}},G_{b_{0}}\right]
^{*}=C_{\;\;a_{0}b_{0}}^{c_{0}}G_{c_{0}},\;\left[ H_{0},G_{a_{0}}\right]
^{*}=V_{\;\;a_{0}}^{b_{0}}G_{b_{0}},  \label{ft3}
\end{equation}
while action (\ref{ft2}) is invariant under the gauge transformations 
\begin{equation}
\delta _{\epsilon }z^{A}=\left[ z^{A},G_{a_{0}}\right] ^{*}\epsilon
^{a_{0}},\;\delta _{\epsilon }u^{a_{0}}=\dot{\epsilon}^{a_{0}}-V_{\;%
\;b_{0}}^{a_{0}}\epsilon ^{b_{0}}-C_{\;\;b_{0}c_{0}}^{a_{0}}\epsilon
^{c_{0}}u^{b_{0}}-Z_{\;\;a_{1}}^{a_{0}}\epsilon ^{a_{1}},  \label{ft4}
\end{equation}
with $Z_{\;\;a_{1}}^{a_{0}}$ the first-stage reducibility functions of the
``free'' theory. In order to generate consistent interactions at the
Hamiltonian level, we deform the action (\ref{ft2}) by adding to it some
interaction terms $S_{0}\rightarrow \tilde{S}_{0}=S_{0}+g\stackrel{(1)}{S}%
_{0}+g^{2}\stackrel{(2)}{S}_{0}+\cdots $, and modify the gauge
transformations (\ref{ft4}) in such a way that the deformed gauge
transformations leave invariant the new action. Consequently, the
deformation of the action (\ref{ft2}) and of the gauge transformations (\ref
{ft4}) produces a deformation of the Hamiltonian ``free'' gauge algebra (\ref
{ft3}) and of the ``free'' reducibility functions. As the BRST charge $%
\Omega _{0}$ and the BRST-invariant Hamiltonian $\stackrel{\left( 0\right) }{%
H}_{B}$ contain all the information on the gauge structure of the ``free''
theory, we can conclude that the deformation of the ``free'' gauge algebra
and ``free'' reducibility functions induces the deformation of the solutions
to the equations $\left[ \Omega _{0},\Omega _{0}\right] ^{*}=0$ and $\left[ 
\stackrel{\left( 0\right) }{H}_{B},\Omega _{0}\right] ^{*}=0$. In
conclusion, the problem of constructing consistent interactions at the
classical Hamiltonian level can be reformulated as a deformation problem of
the BRST charge, respectively, of the BRST-invariant Hamiltonian of the
``free'' theory.

If the interactions are consistently constructed, then the BRST charge of
the ``free'' theory can be deformed as 
\begin{eqnarray}
\Omega _{0}\rightarrow \Omega &=&\Omega _{0}+g\int d^{3}x\,\omega
_{1}+g^{2}\int d^{3}x\,\omega _{2}+O\left( g^{3}\right) =  \nonumber
\label{1} \\
&&\Omega _{0}+g\Omega _{1}+g^{2}\Omega _{2}+O\left( g^{3}\right) ,
\end{eqnarray}
where $\Omega $ should satisfy the equation 
\begin{equation}
\left[ \Omega ,\Omega \right] ^{*}=0.  \label{2}
\end{equation}
Equation (\ref{2}) can be analyzed order by order in the deformation
parameter $g$, leading to 
\begin{equation}
\left[ \Omega _{0},\Omega _{0}\right] ^{*}=0,  \label{3}
\end{equation}
\begin{equation}
2\left[ \Omega _{0},\Omega _{1}\right] ^{*}=0,  \label{4}
\end{equation}
\begin{equation}
2\left[ \Omega _{0},\Omega _{2}\right] ^{*}+\left[ \Omega _{1},\Omega
_{1}\right] ^{*}=0,  \label{5}
\end{equation}
\[
\vdots 
\]
While equation (\ref{3}) is satisfied by assumption, from the remaining
equations we deduce the pieces $\left( \Omega _{k}\right) _{k>0}$ on account
of the ``free'' BRST differential. With the deformed BRST charge at hand, we
then deform the BRST-invariant Hamiltonian of the ``free'' theory 
\begin{eqnarray}
\stackrel{\left( 0\right) }{H}_{B}\rightarrow H_{B} &=&\stackrel{\left(
0\right) }{H}_{B}+g\int d^{3}x\,h_{1}+g^{2}\int d^{3}x\,h_{2}+O\left(
g^{3}\right) =  \nonumber  \label{6} \\
&&\stackrel{\left( 0\right) }{H}_{B}+gH_{1}+g^{2}H_{2}+O\left( g^{3}\right) ,
\end{eqnarray}
and impose that this is precisely the BRST-invariant Hamiltonian of the
deformed system 
\begin{equation}
\left[ H_{B},\Omega \right] ^{*}=0.  \label{7}
\end{equation}
Like in the previous case, equation (\ref{7}) can be decomposed accordingly
the deformation parameter like 
\begin{equation}
\left[ \stackrel{\left( 0\right) }{H}_{B},\Omega _{0}\right] ^{*}=0,
\label{8}
\end{equation}
\begin{equation}
\left[ \stackrel{\left( 0\right) }{H}_{B},\Omega _{1}\right] ^{*}+\left[
H_{1},\Omega _{0}\right] ^{*}=0,  \label{9}
\end{equation}
\begin{equation}
\left[ \stackrel{\left( 0\right) }{H}_{B},\Omega _{2}\right] ^{*}+\left[
H_{1},\Omega _{1}\right] ^{*}+\left[ H_{2},\Omega _{0}\right] ^{*}=0,
\label{10}
\end{equation}
\[
\vdots 
\]
Obviously, equation (\ref{8}) is fulfilled by hypothesis, while from the
other equations one can determine the components $\left( H_{k}\right) _{k>0}$
relying on the BRST symmetry of the ``free'' system. Equations (\ref{3}--\ref
{5}), etc. and (\ref{8}--\ref{10}), etc. stand for the main equations
governing our deformation procedure, and they will be explicitly solved in
the next sections in order to obtain the consistent Hamiltonian interactions
that can be added among a set of two-form gauge fields in four dimensions.

\section{Free BRST differential}

We begin with the Lagrangian action for a set of abelian two-form gauge
fields in first-order form (also known as the abelian Freedman-Townsend
model) 
\begin{equation}
S_{0}^{L}\left[ A_{\mu }^{a},B_{a}^{\mu \nu }\right] =\frac{1}{2}\int
d^{4}x\left( -B_{a}^{\mu \nu }F_{\mu \nu }^{a}+A_{\mu }^{a}A_{a}^{\mu
}\right) ,  \label{11}
\end{equation}
where $B_{a}^{\mu \nu }$ stands for a set of antisymmetric tensor fields,
and the field strength of $A_{\mu }^{a}$ reads as $F_{\mu \nu }^{a}=\partial
_{\mu }A_{\nu }^{a}-\partial _{\nu }A_{\mu }^{a}\equiv \partial _{\left[ \mu
\right. }A_{\left. \nu \right] }^{a}$. After eliminating the second-class
constraints ( the independent `co-ordinates' of the reduced phase-space are $%
A_{i}^{a}$, $B_{a}^{0i}$, $B_{a}^{ij}$ and $\pi _{ij}^{a}$), we remain only
with the first-class ones 
\begin{equation}
G_{i}^{(1)a}\equiv \epsilon _{0ijk}\pi ^{jka}\approx 0,\;G_{i}^{(2)a}\equiv 
\frac{1}{2}\epsilon _{0ijk}F^{jka}\approx 0,  \label{12}
\end{equation}
and the first-class Hamiltonian 
\begin{equation}
H_{0}=\frac{1}{2}\int d^{3}x\left(
B_{a}^{ij}F_{ij}^{a}-A_{i}^{a}A_{a}^{i}+\left( \partial
^{i}B_{0i}^{a}\right) \left( \partial _{j}B_{a}^{0j}\right) \right) \equiv
\int d^{3}x\,h.  \label{13}
\end{equation}
In addition, the functions $G_{i}^{(2)a}$ from (\ref{12}) are first-stage
reducible 
\begin{equation}
\partial ^{i}G_{i}^{(2)b}=0.  \label{14}
\end{equation}
The non-vanishing Dirac brackets among the independent components are
expressed by 
\begin{equation}
\left[ B_{a}^{0i}\left( x\right) ,A_{j}^{b}\left( y\right) \right]
_{x^{0}=y^{0}}^{*}=\delta _{a}^{\;\;b}\delta _{\;\;j}^{i}\delta ^{3}\left( 
{\bf x}-{\bf y}\right) ,  \label{15}
\end{equation}
\begin{equation}
\left[ B_{a}^{ij}\left( x\right) ,\pi _{kl}^{b}\left( y\right) \right]
_{x^{0}=y^{0}}^{*}=\frac{1}{2}\delta _{a}^{\;\;b}\delta _{\;\;k}^{\left[
i\right. }\delta _{\;\;l}^{\left. j\right] }\delta ^{3}\left( {\bf x}-{\bf y}%
\right) ,  \label{16}
\end{equation}
so the Hamiltonian gauge algebra reads as 
\begin{equation}
\left[ G_{i}^{(1)a},G_{j}^{(1)b}\right] ^{*}=0,\;\left[
G_{i}^{(1)a},G_{j}^{(2)b}\right] ^{*}=0,\;\left[
G_{i}^{(2)a},G_{j}^{(2)b}\right] ^{*}=0,  \label{17}
\end{equation}
\begin{equation}
\left[ H_{0},G_{i}^{(1)a}\right] ^{*}=G_{i}^{(2)a},\;\left[
H_{0},G_{i}^{(2)a}\right] ^{*}=0.  \label{18}
\end{equation}
Then, the BRST charge and BRST-invariant Hamiltonian of the free theory are
given by 
\begin{equation}
\Omega _{0}=\int d^{3}x\left( G_{i}^{(1)a}\eta _{a}^{\left( 1\right)
i}+G_{i}^{(2)a}\eta _{a}^{\left( 2\right) i}+\eta _{a}\partial ^{i}{\cal P}%
_{i}^{\left( 2\right) a}\right) ,  \label{19}
\end{equation}
\begin{equation}
\stackrel{\left( 0\right) }{H}_{B}=H_{0}+\int d^{3}x\eta _{a}^{\left(
1\right) i}{\cal P}_{i}^{\left( 2\right) a}.  \label{20}
\end{equation}
In (\ref{19}--\ref{20}), $\eta _{a}^{\left( 1\right) i}$ and $\eta
_{a}^{\left( 2\right) i}$ stand for the fermionic ghost number one ghosts, $%
\eta _{a}$ denote the bosonic ghost number two ghosts for ghosts, while the $%
{\cal P}$ `s represent their corresponding antighosts. The ghost number ($%
{\rm gh}$) is defined like the difference between the pure ghost number ($%
{\rm pgh}$) and the antighost number (${\rm antigh}$), with 
\begin{equation}
{\rm pgh}\left( z^{A}\right) =0,\;{\rm pgh}\left( \eta ^{\Gamma }\right)
=1,\;{\rm pgh}\left( \eta _{a}\right) =2,\;{\rm pgh}\left( {\cal P}_{\Gamma
}\right) =0,\;{\rm pgh}\left( {\cal P}^{a}\right) =0,  \label{21}
\end{equation}
\begin{equation}
{\rm antigh}\left( z^{A}\right) =0,\;{\rm antigh}\left( \eta ^{\Gamma
}\right) =0,\;{\rm antigh}\left( \eta _{a}\right) =0,  \label{22}
\end{equation}
\begin{equation}
{\rm antigh}\left( {\cal P}_{\Gamma }\right) =1,\;{\rm antigh}\left( {\cal P}%
^{a}\right) =2,  \label{23}
\end{equation}
where 
\begin{equation}
z^{A}=\left( A_{i}^{a},B_{a}^{\mu \nu },\pi _{ij}^{a}\right) ,\eta ^{\Gamma
}=\left( \eta _{a}^{\left( 1\right) i},\eta _{a}^{\left( 2\right) i}\right) ,%
{\cal P}_{\Gamma }=\left( {\cal P}_{i}^{\left( 1\right) a},{\cal P}%
_{i}^{\left( 2\right) a}\right) .  \label{24}
\end{equation}
The BRST differential $s\bullet =\left[ \bullet ,\Omega _{0}\right] ^{*}$ of
the free theory splits as 
\begin{equation}
s=\delta +\gamma ,  \label{25}
\end{equation}
where $\delta $ is the Koszul-Tate differential, and $\gamma $ represents
the exterior longitudinal derivative along the gauge orbits. These operators
act like 
\begin{equation}
\delta z^{A}=0,\;\delta \eta ^{\Gamma }=0,\;\delta \eta _{a}=0,  \label{26}
\end{equation}
\begin{equation}
\delta {\cal P}_{i}^{\left( 1\right) a}=-\epsilon _{0ijk}\pi ^{jka},\;\delta 
{\cal P}_{i}^{\left( 2\right) a}=-\frac{1}{2}\epsilon
_{0ijk}F^{jka},\;\delta {\cal P}^{a}=-\partial ^{i}{\cal P}_{i}^{\left(
2\right) a},  \label{27}
\end{equation}
\begin{equation}
\gamma A_{i}^{a}=0,\;\gamma B_{a}^{0i}=\epsilon ^{0ijk}\partial _{j}\eta
_{ka}^{\left( 2\right) },\;\gamma B_{a}^{ij}=\epsilon ^{0ijk}\eta
_{ka}^{\left( 1\right) },\;\gamma \pi _{ij}^{a}=0,  \label{28}
\end{equation}
\begin{equation}
\gamma \eta _{a}^{\left( 1\right) i}=0,\;\gamma \eta _{a}^{\left( 2\right)
i}=\partial ^{i}\eta _{a},\;\gamma \eta _{a}=0,  \label{29}
\end{equation}
\begin{equation}
\gamma {\cal P}_{i}^{\left( 1\right) a}=0,\;\gamma {\cal P}_{i}^{\left(
2\right) a}=0,\;\gamma {\cal P}^{a}=0.  \label{30}
\end{equation}
The above formulas will be used in the next sections at the deformation
procedure.

\section{Deformed BRST charge}

In order to derive the deformed BRST charge resulting from (\ref{19}), we
proceed to solving the equations (\ref{4}--\ref{5}), etc., paying attention
to the fact that the BRST differential of the uncoupled model decomposes
like in (\ref{25}). Equation (\ref{4}) is satisfied if and only if $\omega
_{1}$ is a $s$-co-cycle modulo the exterior spatial derivative $\tilde{d}%
=dx^{i}\partial _{i}$, i.e., it fulfills 
\begin{equation}
s\omega _{1}=\partial _{k}j^{k},  \label{31}
\end{equation}
for some $j^{k}$. For solving the above equation, we develop $\omega _{1}$
accordingly the antighost number 
\begin{equation}
\omega _{1}=\stackrel{\left( 0\right) }{\omega }_{1}+\stackrel{\left(
1\right) }{\omega }_{1}+\cdots +\stackrel{\left( J\right) }{\omega }_{1},\;%
{\rm antigh}\left( \stackrel{\left( I\right) }{\omega }_{1}\right) =I,
\label{32}
\end{equation}
and take into account that the last term in (\ref{32}) can be assumed to be
annihilated by $\gamma $. As ${\rm antigh}\left( \stackrel{\left( J\right) }{%
\omega }_{1}\right) =J$ and ${\rm gh}\left( \stackrel{\left( J\right) }{%
\omega }_{1}\right) =1$, it results that ${\rm pgh}\left( \stackrel{\left(
J\right) }{\omega }_{1}\right) =J+1$. On the other hand, we observe that the
ghosts for ghosts are $\gamma $-invariant, such that we can take $\stackrel{%
\left( J\right) }{\omega }_{1}$ under the form 
\begin{equation}
\stackrel{\left( J\right) }{\omega }_{1}=\alpha ^{a_{1}a_{2}\cdots
a_{N}}\eta _{a_{1}}\eta _{a_{2}}\cdots \eta _{a_{N}},  \label{33}
\end{equation}
where $N$ is a nonnegative integer with $2N=J+1$. This further enforces that 
$J$ should be odd, $J=1,3,5\cdots $. Under this choice, it is simple to
check that the $\gamma $-invariant coefficients $\alpha ^{a_{1}a_{2}\cdots
a_{N}}$ have to pertain to $H_{J}\left( \delta |\tilde{d}\right) $.
Nevertheless, using the results from \cite{31} adapted to the Hamiltonian
case, it follows that $H_{J}\left( \delta |\tilde{d}\right) =0$ for all $J>2$%
, which leads to $J=1$. Consequently, we find that 
\begin{equation}
\omega _{1}=\stackrel{\left( 0\right) }{\omega }_{1}+\stackrel{\left(
1\right) }{\omega }_{1},  \label{34}
\end{equation}
where $\stackrel{\left( 1\right) }{\omega }_{1}=\alpha ^{a}\eta _{a}$, with
the coefficients $\alpha ^{a}$ from $H_{1}\left( \delta |\tilde{d}\right) $,
i.e. 
\begin{equation}
\delta \alpha ^{a}=\partial ^{k}m_{k}^{a},  \label{35}
\end{equation}
for some $m_{k}^{a}$. From (\ref{27}), it results that we have $\alpha
^{a}=\alpha _{\;ib}^{a}{\cal P}^{\left( 2\right) ib}$, such that 
\begin{equation}
\delta \alpha ^{a}=-\frac{1}{2}\alpha _{\;ib}^{a}\epsilon ^{0ijk}F_{jk}^{b}.
\label{36}
\end{equation}
In order to restore a total derivative in the right hand-side of (\ref{36}),
we choose $\alpha _{\;ib}^{a}=f_{\;bc}^{a}A_{i}^{c}$, with $f_{\;bc}^{a}$
some constants antisymmetric in the lower indices, $f_{\;bc}^{a}=-f_{%
\;cb}^{a}$. With this choice, we find that $\delta \alpha ^{a}=\partial
_{j}\left( -\frac{1}{2}f_{\;bc}^{a}\epsilon ^{0ijk}A_{k}^{b}A_{i}^{c}\right) 
$, while 
\begin{equation}
\stackrel{\left( 1\right) }{\omega }_{1}=f_{\;bc}^{a}A_{i}^{c}{\cal P}%
^{\left( 2\right) ib}\eta _{a}.  \label{37}
\end{equation}
Now, we investigate the equation (\ref{31}) at antighost number zero, namely 
\begin{equation}
\delta \stackrel{\left( 1\right) }{\omega }_{1}+\gamma \stackrel{\left(
0\right) }{\omega }_{1}=\partial ^{j}\nu _{j},  \label{38}
\end{equation}
for some $\nu _{j}$. Using (\ref{37}), after some computation we arrive at 
\begin{equation}
\delta \stackrel{\left( 1\right) }{\omega }_{1}=\partial ^{j}\left( -\frac{1%
}{2}f_{\;bc}^{a}\epsilon ^{0ijk}A_{k}^{b}A_{i}^{c}\eta _{a}\right) +\gamma
\left( \frac{1}{2}f_{\;bc}^{a}\epsilon ^{0ijk}A_{k}^{b}A_{i}^{c}\eta
_{ja}^{\left( 2\right) }\right) ,  \label{39}
\end{equation}
which further gives 
\begin{equation}
\stackrel{\left( 0\right) }{\omega }_{1}=-\frac{1}{2}f_{\;bc}^{a}\epsilon
^{0ijk}A_{k}^{b}A_{i}^{c}\eta _{ja}^{\left( 2\right) }.  \label{40}
\end{equation}
In this way, we have completely determined the first-order deformation of
the BRST charge 
\begin{equation}
\Omega _{1}=\int d^{3}x\left( f_{\;bc}^{a}A_{i}^{c}{\cal P}^{\left( 2\right)
ib}\eta _{a}-\frac{1}{2}f_{\;bc}^{a}\epsilon ^{0ijk}A_{j}^{b}A_{k}^{c}\eta
_{ia}^{\left( 2\right) }\right) .  \label{41}
\end{equation}
Next, from (\ref{5}) we conclude that the deformation is consistent also at
order $g^{2}$ if and only if $\left[ \Omega _{1},\Omega _{1}\right] ^{*}$ is 
$s$-exact. In the meantime, with the help of (\ref{41}) we deduce that 
\begin{equation}
\left[ \Omega _{1},\Omega _{1}\right] ^{*}=\frac{1}{3}f_{\;a\left[ d\right.
}^{e}f_{\;\left. bc\right] }^{a}\int d^{3}x\left( \epsilon
^{0ijk}A_{i}^{d}A_{j}^{b}A_{k}^{c}\eta _{e}\right) .  \label{42}
\end{equation}
Thus, the integrand of $\left[ \Omega _{1},\Omega _{1}\right] ^{*}$ cannot
be $s$-exact modulo $\tilde{d}$, so it should vanish. This can be attained
if and only if the constants $f_{\;bc}^{a}$ fulfill the Jacobi identity 
\begin{equation}
f_{\;a\left[ d\right. }^{e}f_{\;\left. bc\right] }^{a}=0,  \label{43}
\end{equation}
hence if and only if they represent the structure constants of a Lie
algebra. Accordingly, we find that $\Omega _{2}=0$. Moreover, the
higher-order equations that govern the deformation of the BRST charge are
satisfied for $\Omega _{3}=\Omega _{4}=\cdots =0$. In conclusion, the
complete deformed BRST charge is expressed precisely by $\Omega =\Omega
_{0}+g\Omega _{1}$.

\section{Deformed BRST-invariant Hamiltonian}

In the sequel we determine the deformed BRST-invariant Hamiltonian
corresponding to (\ref{20}) with the help of the equations (\ref{9}--\ref{10}%
), etc. We begin with the equation (\ref{9}), whose first term is found of
the type 
\begin{eqnarray}
\left[ \stackrel{\left( 0\right) }{H}_{B},\Omega _{1}\right] ^{*} &=&\int
d^{3}xf_{\;bc}^{a}\left( \epsilon ^{0ijk}\left( \frac{1}{2}\eta
_{ia}^{\left( 1\right) }A_{j}^{b}+\eta _{ia}^{\left( 2\right) }\partial
_{j}\left( \partial ^{l}B_{0l}^{b}\right) \right) A_{k}^{c}-\right. 
\nonumber  \label{43a} \\
&&\left. \eta _{a}{\cal P}^{\left( 2\right) ib}\partial _{i}\left( \partial
^{l}B_{0l}^{c}\right) \right) =\int d^{3}x\lambda .
\end{eqnarray}
We can thus write (\ref{9}) in the equivalent form 
\begin{equation}
sh_{1}+\lambda =\partial _{j}\rho ^{j},  \label{44}
\end{equation}
for some $\rho ^{j}$. The solution to (\ref{44}) is expressed by 
\begin{equation}
h_{1}=-f_{\;bc}^{a}\left( \frac{1}{2}A_{j}^{b}A_{k}^{c}B_{a}^{jk}+\left(
\partial ^{l}B_{0l}^{b}\right) \left( B_{a}^{0i}A_{i}^{c}+\eta _{a}^{\left(
2\right) i}{\cal P}_{i}^{\left( 2\right) c}+\eta _{a}{\cal P}^{c}\right)
\right) ,  \label{45}
\end{equation}
which further yields 
\begin{equation}
sh_{1}+\lambda =\partial _{j}\left( f_{\;bc}^{a}\left( \partial
^{l}B_{0l}^{b}\right) \left( \epsilon ^{0ijk}\eta _{ia}^{\left( 2\right)
}A_{k}^{c}+\eta _{a}{\cal P}^{\left( 2\right) jc}\right) \right) .
\label{46}
\end{equation}
On behalf of $h_{1}$, we approach now the equation (\ref{10}). The first
term in (\ref{10}) is equal to zero as $\Omega _{2}=0$, while the second one
is given by 
\begin{eqnarray}
\left[ H_{1},\Omega _{1}\right] ^{*} &=&f_{\;bc}^{a}f_{\;de}^{c}\int
d^{3}x\left( \epsilon ^{0ijl}A_{j}^{e}\eta _{i}^{\left( 2\right) d}-{\cal P}%
^{\left( 2\right) le}\eta ^{d}\right) \times  \nonumber  \label{47} \\
&&\partial _{l}\left( B_{a}^{0k}A_{k}^{b}+\eta _{a}^{\left( 2\right) k}{\cal %
P}_{k}^{\left( 2\right) b}+\eta _{a}{\cal P}^{b}\right) \equiv \int
d^{3}x\beta .
\end{eqnarray}
This means that equation (\ref{10}) can be alternatively written as 
\begin{equation}
sh_{2}+\beta =\partial _{l}\mu ^{l},  \label{48}
\end{equation}
for some $\mu ^{l}$. After some computation, we find that its solution is 
\begin{eqnarray}
&&h_{2}=f_{\;bc}^{a}f_{\;de}^{c}B_{a}^{0k}A_{k}^{b}\left( \frac{1}{2}%
B_{0i}^{d}A^{ie}+\eta ^{\left( 2\right) id}{\cal P}_{i}^{\left( 2\right)
e}+\eta ^{d}{\cal P}^{e}\right) +  \nonumber  \label{49} \\
&&\frac{1}{2}f_{\;bc}^{a}f_{\;de}^{c}\left( \eta _{a}{\cal P}^{b}\eta ^{d}%
{\cal P}^{e}+\eta _{a}^{\left( 2\right) k}{\cal P}_{k}^{\left( 2\right)
b}\eta ^{\left( 2\right) jd}{\cal P}_{j}^{\left( 2\right) e}+2\eta _{a}{\cal %
P}^{b}\eta ^{\left( 2\right) jd}{\cal P}_{j}^{\left( 2\right) e}\right) ,
\end{eqnarray}
where 
\begin{equation}
\mu ^{l}=f_{\;bc}^{a}f_{\;de}^{c}\left( \epsilon ^{0ijl}A_{j}^{e}\eta
_{i}^{\left( 2\right) d}-{\cal P}^{\left( 2\right) le}\eta ^{d}\right)
\left( B_{a}^{0k}A_{k}^{b}+\eta _{a}^{\left( 2\right) k}{\cal P}_{k}^{\left(
2\right) b}+\eta _{a}{\cal P}^{b}\right) .  \label{49a}
\end{equation}
In addition, we derive that $\left[ H_{2},\Omega _{1}\right] =0$. Then, the
equation of order $g^{3}$ associated with the BRST-invariant Hamiltonian is
verified for $h_{3}=0$ as all the terms but that involving $h_{3}$ vanish.
Further, all the higher-order deformation equations are checked if we take $%
H_{4}=H_{5}=\cdots =0$. In consequence, the complete deformed BRST-invariant
Hamiltonian for the model under study reads as $H_{B}=\stackrel{\left(
0\right) }{H}_{B}+g\int d^{3}x\,h_{1}+g^{2}\int d^{3}x\,h_{2}$, where $h_{1}$
and $h_{2}$ are given by (\ref{45}), respectively, (\ref{49}).

\section{Identification of the deformed model}

At this point, we are able to identify the resulting interacting theory.
Synthesizing the results from the previous two sections, so far we solved
the deformation equations associated with the BRST charge and BRST-invariant
Hamiltonian for the free theory, and obtained that their complete consistent
solutions are respectively given by 
\begin{eqnarray}
\Omega &=&\int d^{3}x\left( \frac{1}{2}\epsilon _{0ijk}\left(
F^{jka}-gf_{\;bc}^{a}A^{jb}A^{kc}\right) \eta _{a}^{\left( 2\right)
i}+\right.  \nonumber  \label{50} \\
&&\left. \eta _{a}\left( D^{i}\right) _{\;\;b}^{a}{\cal P}_{i}^{\left(
2\right) b}+\epsilon _{0ijk}\pi ^{jka}\eta _{a}^{\left( 1\right) i}\right) ,
\end{eqnarray}
\begin{eqnarray}
&&H_{B}=\int d^{3}x\left( \frac{1}{2}B_{a}^{ij}\left(
F_{ij}^{a}-gf_{\;bc}^{a}A_{i}^{b}A_{j}^{c}\right) -\frac{1}{2}%
A_{i}^{a}A_{a}^{i}+\eta _{a}^{\left( 1\right) i}{\cal P}_{i}^{\left(
2\right) a}+\right.  \nonumber  \label{51} \\
&&\frac{1}{2}\left( \left( D^{i}\right) _{\;\;b}^{a}B_{0i}^{b}\right) \left(
\left( D_{j}\right) _{a}^{\;\;c}B_{c}^{0j}\right) -gf_{\;bc}^{a}\left( \eta
_{a}^{\left( 2\right) i}{\cal P}_{i}^{\left( 2\right) c}+\eta _{a}{\cal P}%
^{c}\right) \left( D^{l}\right) _{\;\;d}^{b}B_{0l}^{d}+  \nonumber \\
&&\left. \frac{1}{2}g^{2}f_{\;bc}^{a}f_{\;de}^{c}\left( \eta _{a}{\cal P}%
^{b}\eta ^{d}{\cal P}^{e}+\eta _{a}^{\left( 2\right) k}{\cal P}_{k}^{\left(
2\right) b}\eta ^{\left( 2\right) jd}{\cal P}_{j}^{\left( 2\right) e}+2\eta
_{a}{\cal P}^{b}\eta ^{\left( 2\right) jd}{\cal P}_{j}^{\left( 2\right)
e}\right) \right) ,
\end{eqnarray}
where we used the notations $\left( D^{i}\right) _{\;\;b}^{a}=\delta
_{\;\;b}^{a}\partial ^{i}+gf_{\;bc}^{a}A^{ic}$ and $\left( D^{i}\right)
_{b}^{\;\;a}=\delta _{b}^{\;\;a}\partial ^{i}-gf_{\;bc}^{a}A^{ic}$. From the
antighosts-independent component in (\ref{50}) we read that only the latter
set in the initial first-class constraints (\ref{12}) are deformed 
\begin{equation}
\gamma _{i}^{\left( 2\right) a}\equiv \frac{1}{2}\epsilon _{0ijk}\left(
F^{jka}-gf_{\;bc}^{a}A^{jb}A^{kc}\right) \approx 0,  \label{52}
\end{equation}
while the first set is kept unchanged. Another interesting aspect is that
the resulting BRST charge contains no pieces quadratic in the ghost number
one ghosts, hence the gauge algebra (in the Dirac bracket) of the deformed
first-class constraints remains abelian, being not affected by the
deformation method. Moreover, as can be noticed from the term linear in the
ghosts for ghosts, the original reducibility relations (\ref{14}) are also
deformed, the new reducibility relations corresponding to (\ref{52}) being
of the form 
\begin{equation}
\left( D^{i}\right) _{\;\;b}^{a}\gamma _{i}^{\left( 2\right) b}=0.
\label{53}
\end{equation}
Analyzing the structure of the pieces in (\ref{51}) that involve neither
ghosts nor antighosts, we discover that the first-class Hamiltonian of the
deformed theory reads as 
\begin{eqnarray}
H &=&\frac{1}{2}\int d^{3}x\left( B_{a}^{ij}\left(
F_{ij}^{a}-gf_{\;bc}^{a}A_{i}^{b}A_{j}^{c}\right) -A_{i}^{a}A_{a}^{i}+\right.
\nonumber  \label{54} \\
&&\left. \left( \left( D^{i}\right) _{\;\;b}^{a}B_{0i}^{b}\right) \left(
\left( D_{j}\right) _{a}^{\;\;c}B_{c}^{0j}\right) \right) ,
\end{eqnarray}
while from the components linear in the antighost number one antighosts we
find that the Dirac brackets among the new first-class Hamiltonian and
first-class constraint functions $\gamma _{i}^{\left( 2\right) a}$ are
modified as 
\begin{equation}
\left[ H,\gamma _{i}^{\left( 2\right) a}\right] ^{*}=-gf_{\;bc}^{a}\left(
\left( D^{j}\right) _{\;\;d}^{b}B_{0j}^{d}\right) \gamma _{i}^{\left(
2\right) c},  \label{55}
\end{equation}
the others being not altered by the deformation mechanism. The first-class
constraints and first-class Hamiltonian generated until now along the
deformation scheme reveal precisely the consistent Hamiltonian interactions
that can be introduced among a set of two-form gauge fields, which actually
produce the non-abelian Freedman-Townsend model in four-dimensions. As the
first-class constraints generate gauge transformations, we can state that
the added interactions deform the gauge transformations, the reducibility
relations, but not the algebra of gauge transformations (due to the
abelianity of the deformed first-class constraints).

The Lagrangian version corresponding to the deformed model constructed in
the above can be inferred in the usual manner via the extended and total
formalisms, which then lead to the expected Lagrangian action \cite{32} 
\begin{equation}
\tilde{S}_{0}^{L}\left[ A_{\mu }^{a},B_{a}^{\mu \nu }\right] =\frac{1}{2}%
\int d^{4}x\left( -B_{a}^{\mu \nu }H_{\mu \nu }^{a}+A_{\mu }^{a}A_{a}^{\mu
}\right) ,  \label{56}
\end{equation}
that is invariant under the gauge transformations 
\begin{equation}
\delta _{\epsilon }B_{\mu \nu }^{a}=\epsilon _{\mu \nu \lambda \rho }\left(
D^{\lambda }\right) _{\;\;b}^{a}\epsilon ^{\rho b},\;\delta _{\epsilon
}A_{\mu }^{a}=0.  \label{58}
\end{equation}
The notation $H_{\mu \nu }^{a}$ signifies the field strength of Yang-Mills
fields 
\begin{equation}
H_{\mu \nu }^{a}=F_{\mu \nu }^{a}-gf_{\;bc}^{a}A_{\mu }^{b}A_{\nu }^{c}.
\label{57}
\end{equation}
The deformation of the gauge transformations of the two-forms is due exactly
to the term linear in the deformation parameter from the constraint
functions (\ref{52}).

\section{Conclusion}

To conclude with, in this paper we have investigated the consistent
Hamiltonian interactions that can be introduced among a set of two-form
gauge fields in four dimensions. Our analysis is based on the deformation of
both BRST charge and BRST-invariant Hamiltonian of the uncoupled version of
the model under study. Starting with the Hamiltonian BRST symmetry of the
free theory, we infer the first-order deformation of the BRST charge by
expanding the co-cycles accordingly the antighost number, and show that it
is consistent also to higher orders in the deformation parameter. With the
deformed BRST charge at hand, we proceed to deriving the corresponding
deformed BRST-invariant Hamiltonian, which turns out to be at most quadratic
in the coupling constant. In this way, we have generated the Hamiltonian
version of the Freedman-Townsend model. As a result of our procedure, the
added interactions deform the gauge transformations, the reducibility
relations, but not the algebra of gauge transformations.

\section*{Acknowledgment}

This work has been supported by a Romanian National Council for Academic
Scientific Research (CNCSIS) grant.

\section*{Note added}

After completion of this work, we became aware of two more papers, \cite{33}
and \cite{34}, containing relevant Hamiltonian results that can be related
to our approach.

\end{document}